\documentclass[%
 reprint,
 amsmath,amssymb,
 aps,
]{revtex4-2}

\usepackage{graphicx}
\usepackage{dcolumn}
\usepackage{bm}
\newcommand*\diff{\mathop{}\!\mathrm{d}}

\begin{document}

\preprint{APS/123-QED}

\title{Dynamic screening of quasiparticles in WS$_2$ monolayers
}

\author{Stefano Calati}
 \affiliation{Humboldt-Universit\"at zu Berlin, Institut f\"{u}r Chemie, Berlin, Germany.}
 \affiliation{
 Fritz-Haber-Institut der Max-Planck-Gesellschaft, Abt. Physikalische Chemie, Berlin, Germany.}%

\author{Qiuyang Li}%
\affiliation{%
 Department of Chemistry, Columbia University, New York, USA}%


\author{Xiaoyang Zhu}

\affiliation{
 Department of Chemistry, Columbia University, New York, USA
}%

\author{Julia St\"{a}hler}
\affiliation{Humboldt-Universit\"at zu Berlin, Institut f\"ur Chemie, Berlin, Germany.
}%
 \affiliation{
 Fritz-Haber-Institut der Max-Planck-Gesellschaft, Abt. Physikalische Chemie, Berlin, Germany.}%


\date{\today}

\begin{abstract}

We unravel the influence of quasiparticle screening in the non-equilibrium exciton dynamics of monolayer WS$_2$.
We report pump photon energy-dependent exciton blue and red-shifts from time-resolved-reflectance contrast measurements.
Based on a phenomenological model, we isolate the effective impact of excitons and free carriers on the renormalization of the quasi-free particle band gap, exciton binding energy and linewidth broadening.
By this, our work does not only provide a comprehensive picture of the competing phenomena governing the exciton dynamics in WS$_2$ upon photoexcitation, but also demonstrates that exciton and carrier contributions to dynamic screening of the Coulomb interaction differ significantly.
\end{abstract}

\maketitle


Excitons in 2D materials attracted substantial interest in fundamental science  \citep{rev_heinz} and are highly relevant for future technological applications \citep{rev_device}.
Particularly, transition metal dichalcogenides (TMDCs) exhibit, due to their two-dimensional nature, a reduced screening of the Coulomb interaction (CIA), which governs the excitonic properties: binding energies on the order of the 100's of meV \citep{Zhu:2015aa},
the quasi-free particle band gap, and, thus, the resonance energy of the exciton. 
The environment \citep{Hsu:2019aa}, preparation technique \citep{Kaviraj:2019aa}, defect density \citep{Kaviraj:2019aa}, temperature \citep{Huang:2016aa}, and doping \citep{Chernikov:2015aa} can change the screening of the CIA and, therefore, change the optical properties of the material.
When the system is driven out of equilibrium, photoexcited quasiparticles (i.e. excitons or quasi-free carriers) further modify the electronic structure by renormalization \cite{Cun_exc,Sie:r_b,Ruppert:2017aa,heinz:pop_in, Pogna,Steinhoff:2017aa,PhysRevResearch.1.022007,Trovatello:2020aa,D0CP03220D,Li:2022PRB} through \emph{dynamic} screening of the CIA. Due to the complexity of these phenomena, previous studies focused on specific aspects and investigated them in depth. 
However, a comprehensive picture of the mechanisms contributing to the dynamic exciton response in TMDCs was not reported yet, and the detailed impact of different types of photoexcited quasiparticles on the exciton dynamics is still to be understood.
For instance, enhanced screening of free carriers and excitons may reduce the exciton binding energy as well as the quasi-free particle band gap, potentially leading to both, blue- and red-shifts of the exciton resonance.

Although both quasiparticle species act on the screening of the CIA, it is \textit{a priori} not clear whether their impact on binding energy and quasi-free carrier band gap is similarly strong, as carriers and excitons show different degrees of (de-)localization and, thus, polarizability.
This work disentangles the influence of exciton and quasi-free carrier (QFC) screening on the non-equilibrium dynamics of the exciton resonance. 
We employ time-resolved optical pump-probe broadband reflectance contrast (tr-RC) to investigate the pump energy- and fluence-dependent response of the A exciton resonance of a monolayer of WS\textsubscript{2} upon photoexcitation. We selectively create excitons or QFC by photoexciting in and above resonance of the A exciton, respectively. Then, we determine the time-dependent dielectric function of the exciton as introduced previously \citep{Ste2021}, tracking its peak-shift and broadening.
These dynamics are then fitted using a simple model description of the transient quasiparticle populations, showing excellent agreement with the data and providing material parameters in line with the literature. Based on this, we conclude that the linewidth broadening is dominated by QFC-exc and exc-exc scattering. 
QFC screening induces a red-shift of the A exciton due to band gap renormalization (BGR) being the dominant effect, while exciton screening leads to a blue-shift caused by a strong exciton binding energy reduction. 
These findings demonstrate that, despite the fact that both quasiparticle types enhance the screening of the CIA, screening by excitons has a very different effect on the exciton resonance energy than carrier screening. This is likely a consequence of their different polarizability and localization.
The model reported in this work, even though it is restricted to the most basic mechanisms, constitutes the ﬁrst quantitative and comprehensive description of the competing phenomena governing the transient optical properties of WS\textsubscript{2}. This sets a basis for a quantitative comparison of existing and future experimental studies to extract reliable material properties and parameters. Ultimately, this model can be extended to include more complex interactions in TMDC materials.

\begin{figure}
\includegraphics{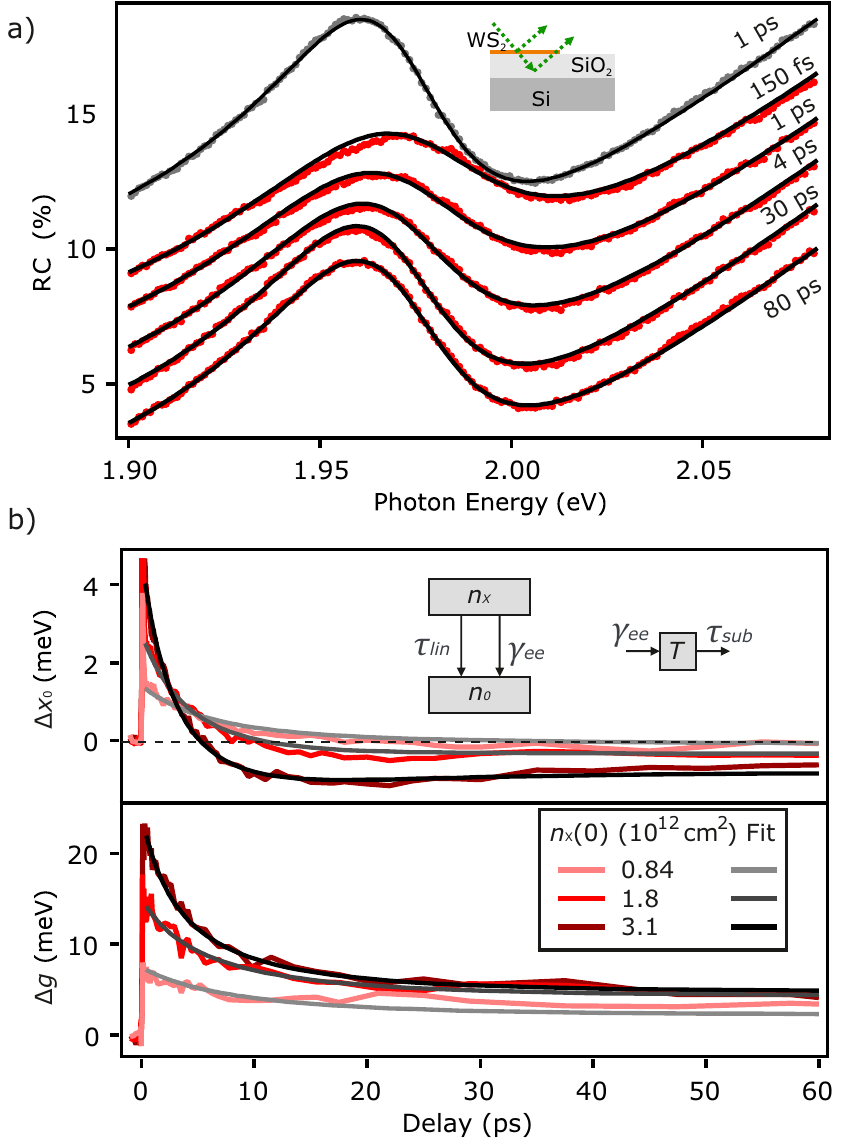}
\caption{\label{fig:1}a) RC spectra (colored markers) for selected delays with fits (grey traces) for an initial exciton density of $n_{x}(0)=\textrm{3.1} \cdot \textrm{10}^{12} \textrm{ cm}^{-2}$. Traces are vertically shifted for clarity. b) Time-dependent exciton peak-shift ($\Delta x_0$) and broadening ($\Delta g$) dynamics for different fluences for resonant pumping. Corresponding fits using Eqs. \ref{eq_shift_r} and \ref{eq_gamma_r}. Inset: Sketch of the two-level model.}
\end{figure}

We study a monolayer of WS\textsubscript{2} on a silicon wafer covered with a thin layer (285 nm) of SiO\textsubscript{2} as sketched in the inset of Fig. \ref{fig:1}.a) using time-resolved reflectance contrast (tr-RC). We chose the commonly used \citep{Cun_exc,Ruppert:2017aa,Chernikov:2015aa} Si/SiO$_2$ as a substrate in order to ensure comparability of our results to previous data. Furthermore, it guarantees a high contrast between the monolayer and the bare substrate. A first pump laser pulse at 1.98 eV or 3.1 eV perturbs the system, and a broadband white light probe pulse \citep{Daniel} monitors the reflectance contrast $RC=(R_{sub}-R_{sub+WS_2})/R_{sub}$ as a function of the delay $t$ between pump and probe pulses.
The experimental details are described in the Supplemental Material (SM).

Fig. \ref{fig:1}.a) shows RC spectra for a few selected pump-probe delays for resonant excitation of the A exciton (1.98 eV). The negative time delay ($t$ = -1 ps, grey) RC spectrum is the steady-state spectrum.
The A exciton signature undergoes peak-shift and broadening upon photoexcitation.
We model the RC spectrum with a Lorentz oscillator and the Fresnel transfer matrix formalism \citep{Stenzel} to account for reflections at the multiple interfaces of the sample, as presented in a previous publication \citep{Ste2021}.
Fits of each spectrum are displayed in Fig. \ref{fig:1}.a), excellently agreeing with the data.
With this approach, we extract the time-dependent exciton peak-shift $\Delta x_0=x_0(t)-x_0(eq)$, linewidth broadening $\Delta g=g(t)-g(eq)$, and spectral weight change at different pump fluences, as displayed in Fig. \ref{fig:1}.b).
We observe an abrupt blue-shift of the resonance upon photoexcitation. The blue-shift ($\Delta x_0 >0$) turns into a red-shift ($\Delta x_0 <0$) at larger delays; the sooner, the higher the pump fluence (cf. crossing of the traces with the dashed zero line). Furthermore, we observe a broadening persisting for tens of picoseconds after photoexcitation (bottom panel). 
The spectral weight does not change within the noise level of our measurements (c.f. SM).

Note that when the pump photon energy is tuned into resonance with the exciton, only excitons are photoexcited initially \citep{Steinhoff:2017aa}.
Based on this, we describe the A exciton dynamics in the simplest way, assuming a phenomenological two-level model sketched in the inset of Fig. \ref{fig:1}.b). The pump generates an initial photoexcited exciton density $n_x(0)=n_{h\nu}$, (with $n_{h\nu}$ the number of absorbed photons) calculated from the pump fluence and spectrum as described in \citep{Ste2021} and in the SM.
The excitons relax to the ground state $n_{0}$ through a linear recombination channel \citep{Yuan:2015aa} (with a rate 1/$\tau_{lin}$) and a quadratic term that is likely attributed to Auger recombination \citep{Sun:2014aa} with the rate $\gamma_{e-e}$.
The time-dependent exciton population density $n_x$ is retrieved by solving:
\begin{equation}
\frac{\diff{} n_x}{\diff{} t}=-\frac{1}{\tau_{lin}} n_x- \gamma_{e-e} n_x^2
\label{eq_r_x}
\end{equation}

Importantly, the Auger recombination induces energy transfer from the electronic to the phonon system, ultimately increasing the lattice temperature and resulting in a red-shift of the exciton \citep{Ruppert:2017aa}.
We assume that the lattice temperature increase is proportional to $\gamma_{e-e} n_{x}^2$, the rate of exciton Auger recombination times the square of the exciton density, as all non-radiatively dissipated energy eventually ends up in the phonon bath. This is sketched in the inset of Fig. \ref{fig:1}.b). 
The excess thermal energy of the lattice is dissipated to the substrate with a rate 1/$\tau_{sub}$.
The temperature evolution $T(t)$ is given by:
\begin{equation}
\frac{\diff{} T(t)}{\diff{}t}=+\gamma_{e-e} n_{x}^2\frac{x_{0}(eq)}{\rho c_p d}-\frac{1}{\tau_{sub}}T
\label{eq_r_T}
\end{equation}
with the bulk WS\textsubscript{2} density \mbox{$\rho=\textrm{7.5 g/cm}^3$} \citep{Ruppert:2017aa}, the specific heat \mbox{$c_p=\textrm{0.25 J/gK}$}  \citep{Ruppert:2017aa}, and the height of the monolayer $d= \textrm{0.7 nm}$ measured with AFM (c.f. SM).

We now relate the exciton peak-shift $\Delta x_0(t)$ and broadening $\Delta g(t)$ to the time-dependent lattice temperature increase and the exciton density $n_x$. In a first approach, it can be assumed that the broadening is proportional to the scattering events between excitons, corresponding to a collisional broadening picture. The broadening is linear in the density of photoexcited quasiparticles \citep{Moody:2015aa}, leading to

\begin{equation}
\Delta g(t)= n_x D
\label{eq_gamma_r}
\end{equation}
with an excitation-induced broadening parameter $D$.

For the description of the exciton peak-shift, we assume the binding energy to be equal to the work needed to bring the electron to an infinite distance from the hole in a screened environment, given by the formula:
\begin{equation}
E_B(t)=\frac{e^2}{4\pi r_0 \varepsilon_0(\varepsilon_{r_{s}}+\varepsilon_{r_{dyn}}(t))}
\label{eq_BE}
\end{equation}
where $e$ is the elementary charge, $\varepsilon_0 $ is the vacuum dielectric constant, and $r_0$ is the distance between the electron and the hole forming the exciton, approximated to the Bohr radius of 1 nm \citep{heinz:pop_in}.
The relative permittivity consists of a static term $\varepsilon_{r_{s}}$ attributed to the effective screening of the dielectric environment. We fix it to the average of the relative permittivity of SiO\textsubscript{2} ($\varepsilon_{r} = 3.9$) and air $\varepsilon_{r_{s}}=2.45$. The second, dynamic term $\varepsilon_{r_{dyn}}(t)=\alpha n_x(t)$ is proportional to the exciton density through the coefficient $\alpha$ and describes the dynamic screening of other excitons. 
The exciton resonance energy is then given by the following expression:
\begin{equation}
\Delta x_0(t)= E_{B}(eq)-E_B(n_x(t))+\beta\Delta T(t)
\label{eq_shift_r}
\end{equation}
Here, the last term describes the temperature-induced red-shift based on Eq. \ref{eq_r_T} and $E_b(eq)$ is given by Eq. \ref{eq_BE} for $\varepsilon_{r_{dyn}}=0$.

We now test the accuracy of this model by performing a global fit to the fluence-dependent peak-shift and broadening with the functions Eq. \ref{eq_shift_r} and Eq. \ref{eq_gamma_r}, respectively. 
We start by fitting the experimental curves 150 fs after time zero, where we can assume an incoherent exciton population \citep{Selig:2018aa}.
The dynamics during the presence of the pump pulse are not analyzed to avoid resonant processes such as the optical Stark effect \citep{Cun_stark,Sie_stark}.
The fit results are shown in Fig. \ref{fig:1}.b) 
The excellent agreement between the fit and the extracted peak-shift and broadening confirms that this simple model can reproduce the exciton dynamics for excitation densities below the Mott transition.
This includes the reproduction of the initial blue-shift by the exciton density-dependent binding energy, the following red-shift due to the temperature rise, and, remarkably, the broadening that is simply proportional to $n_{x}$. 

\begin{figure*}
\includegraphics{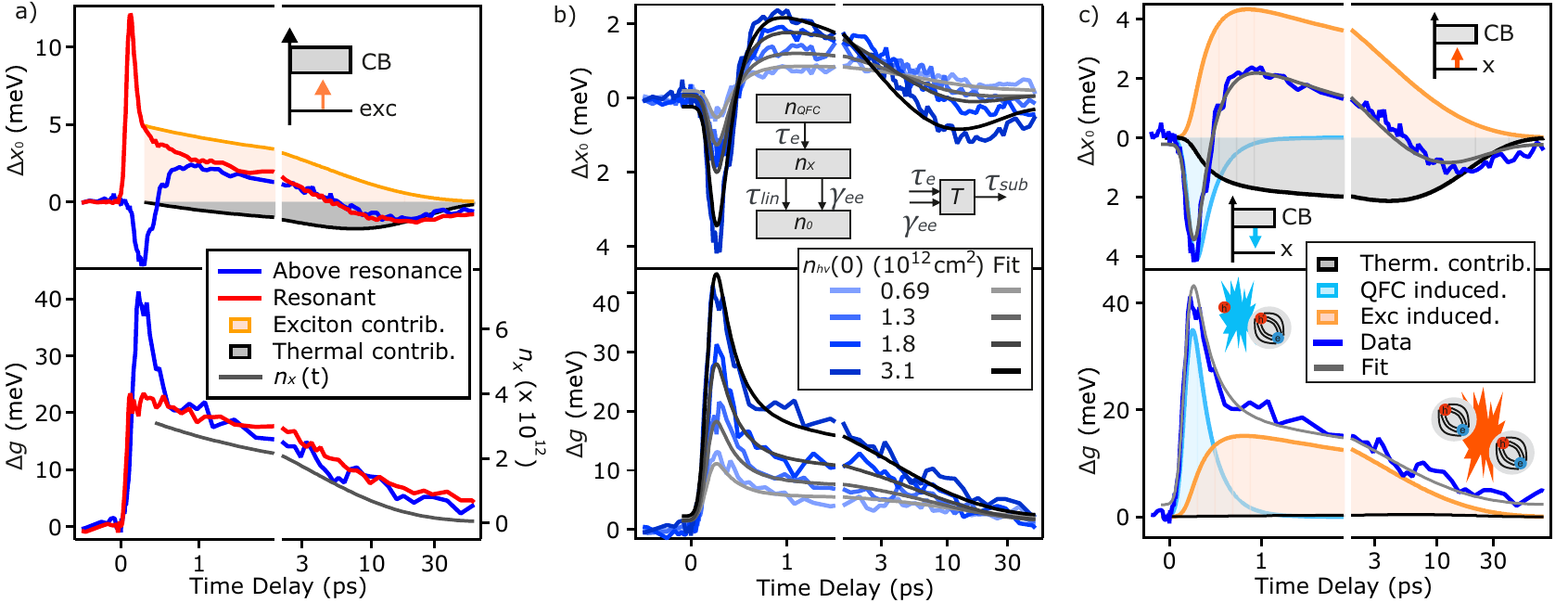}
\caption{\label{fig:2}a) Time-dependent exciton peak-shift ($\Delta x_0$ ) and broadening ($\Delta g$) for above-resonance (blue) and resonant (red) pumping with an absorbed photon density $n_{h\nu}(0)=\textrm{3.1} \cdot \textrm{10}^{12} \textrm{ cm}^{-2}$. binding energy reduction (orange solid) and temperature-induced (solid grey) contributions to resonant peak-shift dynamics. The time-dependent exciton density $n_x(t)$ is displayed on the right axis of the bottom panel.
b) Above-resonance pumping: time-dependent exciton peak-shift and broadening for different fluences with corresponding fit using Eq. \ref{eq_nr_excpos} and Eq. \ref{eq_nr_gamma}. Inset: sketch of the three-level model. c) Top panel: Time-dependent exciton peak-shift for above-resonance pumping and corresponding fit with Eq. \ref{eq_nr_excpos} (from Fig. \ref{fig:2}.b)). Time-dependent contributions to the peak-shift originating from temperature increase (grey), QFC (light blue) and excitons (orange).
Bottom panel: photoinduced broadening and respective fit with Eq. \ref{eq_nr_gamma} from Fig. \ref{fig:2}.b). Contribution to the broadening from QFC-exc (light blue) and exc-exc scattering (orange).}
\end{figure*}

Based on this excellent agreement, we deduce that the restriction to the essential mechanisms and simple population considerations are sufficient to describe all aspects of the photoinduced dynamics in monolayer WS\textsubscript{2} upon resonant photoexcitation. These mechanisms are (i) exc-exc scattering as the primary source of the photoinduced broadening, as shown by the similar trend of broadening and $n_x(t)$ in the bottom panel of Fig. \ref{fig:2}.a) (ii) dynamic screening of excitons inducing binding energy reduction and, thus, the blue-shift at early time delays, and (iii) the temperature increase due to the Auger recombination leading to the observed red-shift at later time delays. The relative contributions of exciton screening and temperature-induced shift are displayed in Fig. \ref{fig:2}.a) (top).
Their competition determines the absolute shift of the exciton. For low initial excited quasiparticle densities, the thermally induced red-shift is minimal due to the fact that excitons recombine prevalently through the linear recombination channel. Instead, the higher the initial photoexcitation density, the higher is the fraction of excitons recombining by Auger recombination, thus leading to an increase in temperature.

To further test the robustness of this model, the same analysis was performed on transmittance contrast data taken on a monolayer WS\textsubscript{2} placed on a fused silica substrate, finding that the model accurately reproduces the dynamics also in this independent experiment (c.f. SM). 
The model, thus, describes the effects of excitons on the optical observables independent of sample, measurement technique, and substrate.

Before discussing the extracted fit parameters, we focus on the role of the QFC.
We perform the same tr-RC experiment, but using a pump of 3.1 eV photon energy. This initially generates quasi-free electrons and holes in the system instead of excitons as in the resonant pumping scheme. It is expected that, after they have released their excess energy by scattering processes, they form bound electron-hole pairs. \cite{osti_10017156} Any difference that is observed in the data of this experiment compared to the resonant pumping case will, thus, be a result of the QFC in the system.
As above, we extract the time-dependent $\Delta x_0$ and $\Delta g$ from the tr-RC using the Fresnel transfer matrix formalism \citep{Ste2021,Stenzel}.

Fig. \ref{fig:2}.a) compares exciton peak-shift and broadening for resonant (red) and above-resonance (blue) pumping with the same initial absorbed photon density. The data clearly shows differing dynamics during the first picosecond after photoexcitation. An initial blue- and red-shift is observed for the resonant and above-resonance pumping, respectively. Additionally, a stronger broadening of the exciton resonance is present for above-resonance pumping. The dynamics at later times overlap across the whole delay range after the above-resonance pumping red-shift turns rapidly into a blue-shift, and the broadening relaxes quickly to values similar to the resonant pumping case. 
Qualitatively, the same is observed for the other excitation densities (Fig. \ref{fig:2}.b)).

The initially differing dynamics must result from the initial QFC population that subsequently decays into an exciton population with the corresponding equilibration dynamics.
To quantify this, we extend the phenomenological model by introducing a third level for the QFC population, as depicted in the inset of Fig. \ref{fig:2}.b). The initial QFC density $n_{QFC}(0)=2n_{h\nu}$ relaxes by forming excitons at a rate of 1/$\tau_e$;
\begin{equation}
\frac{\diff{} n_{QFC}(t)}{\diff{} t}= -\frac{1}{\tau_e}n_{QFC}
\label{eq_nr_ncb}
\end{equation}
Due to the quick nature of the exciton formation process, the exponential decay resulting from solving this rate equation is convoluted with a Gaussian to account for the laser pulses' cross correlation. This gives an analytical expression of $n_{QFC}(t)$.
The time-dependent exciton population $n_{x}(t)$ becomes:
\begin{equation}
\frac{\diff{} n_{x}(t)}{\diff{}t}= +\frac{1}{\tau_e}\frac{n_{QFC}}{2}-\frac{1}{\tau_{lin}} n_x- \gamma_{e-e} n_x^2
\label{eq_nr_nx}
\end{equation}
where the first term is the exciton population build-up taking into account that an exciton is formed by two QFC. The remaining terms are analogous to Eq. \ref{eq_r_x}.

The relaxation of hot QFC forming excitons also yields an additional temperature rise \citep{Caruso:2021aa}. We approximate that the QFC excess energy $h\nu_{pump}-x_{0}(eq)$ is transferred to the lattice at the same rate $1/\tau_e$ as the excitons are being formed. Inserting this into Eq. \ref{eq_r_T}, we obtain:
\begin{equation}
\frac{\diff{} T(t)}{\diff{} t}=\frac{1}{\tau_e} n_{h\nu}\frac{h\nu_{pump}-x_{0}(eq)}{\rho c_p d}+\gamma_{e-e} n_{x}^2\frac{x_{0}(eq)}{\rho c_p d}-\frac{1}{\tau_{sub}}T
\label{eq_nr_T}
\end{equation}
We solve Eqs. \ref{eq_nr_nx} and \ref{eq_nr_T} numerically, thus providing the time-dependent exciton density and lattice temperature.

To relate the three-level model to the observed broadening, another broadening channel must be added to Eq. \ref{eq_gamma_r}, corresponding to the QFC-exc scattering:
\begin{equation}
\Delta g(t)= n_xD+n_{QFC}K
\label{eq_nr_gamma}
\end{equation}
Here, $K$ is the excitation-induced broadening parameter for QFC-exc scattering. 

To model the QFC-induced peak-shift, we assume that the QFC effectively induce a red-shift due to BGR caused by enhanced screening, in line with the clear experimental result that the resonance initially red-shifts when QFC are present. We assume this red-shift to be linear in $n_{QFC}$ within the limited fluence range of this work:

\begin{equation}
\Delta x_0(t)= E_B(eq)-E_B(n_x(t))+\beta\Delta T(t) -E_{QFC}n_{QFC}
\label{eq_nr_excpos}
\end{equation}

We globally fit the fluence-dependent exciton position and broadening for the above-resonance pumping with Eqs. \ref{eq_nr_excpos} and \ref{eq_nr_gamma}, keeping the shared parameters extracted from the resonant pumping data fixed. 
The fits are shown in Fig. \ref{fig:2}.b), displaying an excellent agreement with the complex dynamics.
Despite being a simplification, this model captures the global red-shift of the exciton resonance caused by BGR through QFC screening, and the broadening of the exciton resonance through QFC-exc scattering. Note that our model reproduces the linewidth broadening without including a thermal contribution (cf. Eq. \ref{eq_nr_gamma}), different compared to previous works \citep{Selig:2016aa,Ruppert:2017aa}.

We now turn to the parameters extracted from the global fitting, reported in table \ref{tbl1}.
\begin{table}
\begin{ruledtabular}
\begin{tabular}{c c c} 
 $\tau_e$ (fs)& 200$\pm$ 10  &  $\leq$ 1000 \citep{Steinleitner:2017aa} \\
 $\tau_{lin} $ (ps) & 18 $\pm$ 5 & \\
 $\gamma_{e-e}$ (cm$^2$/s) & 0.06 $\pm$ 0.01 & 0.05-0.1  \citep{eeA:0_1,Cunningham:2016aa,Han:2019aa,Fu:19} \\
 $\tau_{sub}$ (ps) & 14 $\pm$ 8 & ps-$\mu$s \citep{Ruppert:2017aa,Liu:2020aa,Zhao:2013aa}\\
 $\beta$ (meV/K)& 0.4 $\pm$ 0.1 & 0.25-0.5 \citep{Ruppert:2017aa} \\ [0.5ex] 
 $\alpha$ (cm$^2$) & (6.6 $\pm$ 0.6) $\cdot$ 10$^{-15}$ &\\
 $E_{QFC}$ (meV cm$^2$)& (1.2$ \pm$ 0.1) $\cdot$ 10$^{-12}$ &\\
 $D$ (meV cm$^2$)& (5.6 $\pm$ 0.2) $\cdot$ 10$^{-12}$ & 5.4 $\cdot$ 10$^{-12}$ \citep{Moody:2015aa}\\
 $K$ (meV cm$^2$) & (9.8 $\pm$ 0.2) $\cdot$ 10$^{-12}$ &\\ 
\end{tabular}
\end{ruledtabular}
 \caption{\label{tbl1}Best-fit parameters of the model.}
\end{table}
The timescales of the relaxation processes occurring after photoexcitation: exciton formation time $\tau_e$, Auger recombination rate $\gamma_{e-e}$, and thermal energy transfer to the substrate $\tau_{sub}$ are consistent with values previously reported  \citep{Steinleitner:2017aa},\citep{eeA:0_1,Cunningham:2016aa,Han:2019aa,Fu:19}\citep{Ruppert:2017aa}, even specifying the values within the ranges reported in literature.
Notably, the expression for the transient temperatures (Eqs. \ref{eq_r_T} and \ref{eq_nr_T}) were able to reproduce the slower dynamics without having to include temperature changes due to the linear exciton recombination channel ($\tau_{lin}$). 
This suggests radiative electron-hole recombination, trapping, or dark exciton formation that do not transfer significant amounts of energy to the lattice. Reported values for radiative recombination vary quite drastically from sub-ps to nanoseconds \citep{Wang:2015aa,Moody:16,Poellmann:2015aa} and also defect trapping of the excitons, as predicted \citep{Wang:2015aa} and measured \citep{Cui:2014aa} previously for MoS\textsubscript{2} and WSe\textsubscript{2} respectively, as well as dark exciton formation \citep{Poellmann:2015aa} for WSe\textsubscript{2} occur on timescales consistent with $\tau_{lin}$. 

The temperature-induced shift coefficient $\beta$, lies within the values reported in literature \citep{Ruppert:2017aa,Liu:2020aa,Zhao:2013aa}.
The excitation-induced broadening parameter for exc-exc scattering $D$ is also consistent with the literature value \citep{Moody:2015aa}, and the one for QFC-exc scattering $K$ is expected to be higher than $D$ due to higher mobility and lower effective mass of QFC \cite{Honold:1989aa}, in agreement with our observation. The consistency of all these parameters with the literature values shows that the model is based on appropriate assumptions and that the most essential mechanisms influencing the exciton resonance dynamics are an accurate description of the dominating processes.

The remaining  two parameters, $E_{QFC}$ and $\alpha$, were not reported in the literature before. While the former determines the red-shift of the exciton resonance induced by QFC screening (cf. Eq.~\ref{eq_nr_excpos}), the latter causes the blue-shift induced by exciton screening that enhances the relative permittivity (cf. Eq.~\ref{eq_BE}). In order to quantitatively compare the QFC- and exciton-induced shifts, we simplify the exciton-induced binding energy reduction to $\Delta E(t)=E_X n_x(t)$, which is possible, as $\varepsilon_{dyn}(t)<<\varepsilon_s$. Here, the proportionality factor $E_X=(E_B(eq) \alpha/\varepsilon_s)=\textrm{1.6} \cdot \textrm{10}^{12} \textrm{ meV/cm}^{2}$ expresses the net exciton screening contribution to the binding energy reduction. It can be directly compared to the value of $E_{QFC}\textrm{=1.2} \cdot \textrm{10}^{12} \textrm{ meV/cm}^2$, which quantifies the effective red-shift induced by quasi-free electron-hole pairs. 

While both, exciton- and QFC-shifts, are a result of the competing processes of binding energy reduction and BGR through screening of the two quasiparticle types, they have \emph{opposite} signs - despite the fact that excitons are composed of electron-hole pairs, as the QFC population. The origin of this opposing screening effect, must, therefore, lie in the different screening of bound \emph{versus} unbound electron-hole pairs. It seems highly likely that the very different degree of localization of the bound electron-hole pairs that extend across a few lattice sites and the delocalized character of electrons and holes in the quasi-free carrier bands is causing this phenomenon. Presumably, the spatially well-separated excitons simply have a negligible impact on the self-energy of the system and do not cause any relevant BGR. The QFC, conversely, affect the band gap significantly, as generally observed in semiconductors, and this effect is exceeding any carrier-induced binding energy reduction, leading to an effective red-shift.

The opposite sign and a comparison of the magnitude of the respective resonance shifts of carriers and excitons is illustrated in Fig. \ref{fig:2}.c) The figure depicts the time-dependent exciton peak-shift (top) and broadening (bottom), and respective fits, for an above-resonance pumping tr-RC dataset with an initial absorbed photon density of $n_{h\nu}(0)=\textrm{3.1 $\cdot$ 10}^{12} \textrm{ cm}^{-2}$.  The QFC and exciton contributions to shift and broadening are displayed in light blue and orange, respectively. Initially, the red-shift ($\Delta x < 0$, light blue) due to QFC dominates and decays quickly with the exciton formation time $\tau_e=200$~fs. The resulting build-up of exciton population then causes an increasing blue-shift ($\Delta x > 0$, orange). At the same time, the temperature-induced BGR ($\Delta x < 0$, black) rises, as the relaxation of carriers heats up the lattice. After roughly half a picosecond, the sum of all these shifts $\Delta x$ (dark blue) equals zero: More than half of the carrier population has turned into excitons.
 
The similar magnitude of QFC- and exciton screening-induced blue- and red-shifts is likely the cause for the varying observations in the literature \cite{Cun_exc,Sie:r_b,Ruppert:2017aa,heinz:pop_in, Pogna,Steinhoff:2017aa,PhysRevResearch.1.022007,Trovatello:2020aa,D0CP03220D,Li:2022PRB}, as excitation density and photon energy have significant impact on this fragile balance. We hope that our simple model will prove useful in interpreting and understanding of previous and future experimental results on the dynamics in WS\textsubscript{2} as well as other two-dimensional systems with reduced static screening.

In summary, the pump fluence- and photon-energy-dependent quasiparticle dynamics of monolayer WS\textsubscript{2} were studied by means of time-resolved reflectance contrast. 
Excitons and quasi-free carriers were selectively photoexcited in the system employing resonant/above-resonance optical pumping.
We showed that a very reduced model describing QFC and exciton population dynamics is able to reproduce the dynamics of the exciton resonance upon photoexcitation fully, including several cross-over regions, while providing parameters in excellent agreement with the literature. Based on this, the photoexcited quasiparticles' effective impact on the exciton dynamics is unveiled. QFC induce broadening of the exciton resonance through QFC-exc scattering and a red-shift of the resonance by BGR. Excitons, instead, induce broadening of the resonance through exc-exc scattering and a blue-shift modelled by binding energy reduction. Remarkably, our model quantifies the screening-induced contributions of carriers and excitons to the overall shift of the exciton resonance, which are very similar, but with opposite signs, likely due to the different degree of localization of the two quasiparticle species. 
This work provides a basic, but still accurate and comprehensive picture of the competing phenomena that govern the exciton dynamics in WS\textsubscript{2} upon photoexcitation, guaranteeing a common ground for comparing experimental studies of TMDCs and the extraction of reliable material properties and parameters. Furthermore, the two/three-level model may serve as a base for possible extensions to include more complex interactions in TMDC materials, maybe even at higher excitation densities above the Mott limit. 

\begin{acknowledgments}

-This work was funded by the Deutsche Forschungsgemeinschaft (DFG, German Research Foundation)—Project-ID 182087777—SFB 951.
X.Y.Z. acknowledges support for sample preparation by the Materials Science and Engineering Research Center (MRSEC) through NSF grant DMR- 2011738.
We thank S. Palato and M. Kapitzke for providing AFM images of our samples.
\end{acknowledgments}

\appendix

\bibliography{apssamp}

\end{document}


\maketitle

		\pagebreak 
\renewbibmacro{in:}{}

\newrefcontext[labelprefix=S]

\section{Methods}

\begin{figure} [ht]
	\centering
		\includegraphics{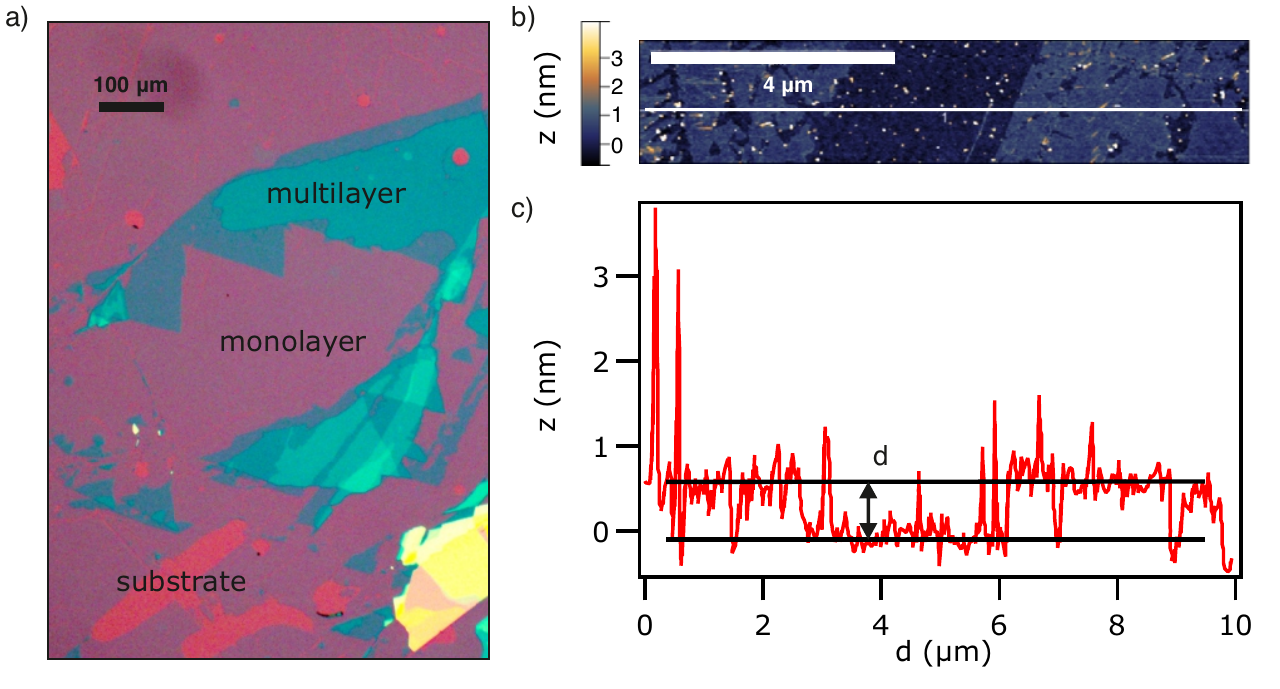}
		\caption{\label{fig_s1}a) Optical microscope images of WS$_2$ samples deposited on Si/SiO$_2$. Hundreds by hundreds of micrometers areas of monolayer are identified as well as substrate and multilayer areas. b)AFM image of a monolayer edge. c) Height cut along the white line on the AFM image. Thickness of the layer reported with the black arrow.
}
\end{figure}

The WS$_2$ monolayers are produced by Au-assisted mechanical exfoliation of a CVD grown large area crystal, as described in detail in Ref  \cite{PhysRevLett.122.246803,Desai:2016}. The monolayer is placed on a silicon wafer covered with a nominal 285 nm layer of silicon oxide (Si/SiO$_2$). Large clean monolayer areas of hundreds of micrometers are present on the sample, as displayed in Fig. \ref{fig_s1}. Places with the clean bare substrate are also present on the sample (not shown in the image). Areas with multilayers are also corresponding to the blue parts in the image. The thickness of the monolayer was measured using a height profile cut, shown in Fig \ref{fig_s1}.c), of an AFM scan of the sample, also reported in Fig \ref{fig_s1}.b), and found to be 0.7 nm.

An optical pump and probe technique is used to measure the time dependence of the reflectivity contrast (RC), defined as $RC= (R_{sub}-R_{sub+WS_2})/R_{sub}$. $R_{sub}$ and $R_{sub+WS_2}$ are the reflectance of the bare substrate and the WS$_2$ monolayer on the substrate, respectively. With this technique, an optical pump pulse brings the system out of equilibrium. After a variable time delay, a second broadband white light pulse probes the reflectance of the monolayer WS$_2$.

The optical pulses at 1.55 eV photon energy are obtained with a Ti:Sa laser system (Coherent RegA), with a 200 kHz repetition rate, and $< $ 40 fs time duration. An optical parametric amplifier (OPA) is used to generate the pump in the case of resonant pumping. The OPA uses a fraction of the laser system’s output to produce 2.0 eV photon energy pulses, which are then compressed to a time duration $<$100 fs. The pump photon energy for the above resonance pumping is 3.1 eV, obtained by the second harmonic generation of the 1.55 eV in a BBO crystal.
The pump pulse is incident on the sample at a 45° angle in both measurements.

The white-light probe, covering a spectral area ranging from 1.85 eV to 2.25 eV,  was compressed and characterized as described in \cite{Daniel}, to 15 fs duration. The s-polarized probe pulse is incident at 45°. 
 
The spectrally resolved reflectance from the sample was measured with a Shamrock 303i spectrometer. The entire reflectance spectrum of the WS$_2$ or substrate was obtained at each time delay. The reflectance of the substrate was measured after every time-resolved scan.

All measurements are performed at room temperature and ambient conditions.

\section{Fit function of tr-RC}

The fit function used to reproduce the RC spectra  in Fig 1.a) is given by the following formula:
\begin{equation}
RC=\frac{R_{sub}(d_{SiO_2},\varphi)-R_{WS_2}(\varepsilon_{WS_2}(x_0,g,I,\varepsilon_{\infty}),\varphi,d_{SiO_2},d_{WS_2})}{R_{sub}}+A+B(x-x_p)+C(x-x_p)^2 
\tag{S.1}\label{Eq-RC}
\end{equation}

$R_{sub}$  is the reflected fraction from the bare substrate and $R_{WS_2}$ the reflected fraction from the monolayer WS$_2$ placed on the substrate. These two quantities are calculated assuming that the exciton contribution to the complex dielectric function of the WS$_2$ can be modeled with a Lorentzian function, given by the expression:

\begin{equation}
\varepsilon_{WS_2}=\varepsilon_{\infty}+\frac{I}{x_{0}^2-x^2-igx}
\tag{S.2}
\label{lor}
\end{equation}

The reflected fraction in both cases is computed using a Fresnel transfer matrix formalism based on the characteristic matrix, described in detail in \cite{Ste:21}. The characteristic matrix expression for a single layer is given by:

\begin{equation}
\hat{\textbf{M}}(z)=\begin{pmatrix}
\cos(k_0\hat{n}z\cos\psi) & -\frac{i}{\hat{n}\cos\psi}\sin(k_0\hat{n}z\cos\psi) \\
-i\hat{n}\cos\psi\sin(k_0\hat{n}z\cos\psi)  & \cos(k_0\hat{n}z\cos\psi) 
\end{pmatrix}
\tag{S.3}
\end{equation}
The characteristic matrix of the stack is given by the multiplication of all the matrixes of the constituent layers.
From this, the reflected fraction is calculated according to the formula:

\begin{equation}
R=\vert r \vert^2=\left| \frac{(m_{11}+m_{12}\hat{n}_{sub}\cos\varphi_{sub})\cos\varphi-(m_{21}+m_{22}\hat{n}_{sub}\cos\varphi_{sub})}{(m_{11}+m_{12}\hat{n}_{sub}\cos\varphi_{sub})\cos\varphi+m_{21}+m_{22}\hat{n}_{sub}\cos\varphi_{sub}} \right|^2
\tag{S.4}
\end{equation}

The polynomial function used in Eq. \ref{Eq-RC} is used to model the background absorption originating from higher-lying electronic transitions.
The thickness of the WS$_2$ monolayer is $d_{WS_2}= 0.7$ nm was measured with AFM as described above. The incidence angle of the probe beam $\varphi=0.77$ rad, and the thickness of the SiO$_2$ layer $d_{SiO_2}=305\pm 20$ nm have been determined from a global fit to various static spectra.
Ultimately, $\varepsilon_{\inf}=16.62$ has been fixed to the value extracted in our previous work \cite{Ste:21} and the complex refractive index of the fused silica and silicon are taken from literature \cite{fs_epsilon,PhysRevB.27.985}

The fit of the steady-state spectrum is performed with Eq.\ref{Eq-RC} which depends on the three parameters of the Lorentzian resonance: central energy, width, spectral weight, and the four parameters of the polynomial. The polynomial background is then kept fixed for the fit of the tr-RC at every time delay.

\section{Time-dependent spectral weight}

	\begin{figure}[ht]
	\centering
	\includegraphics[width=130mm]{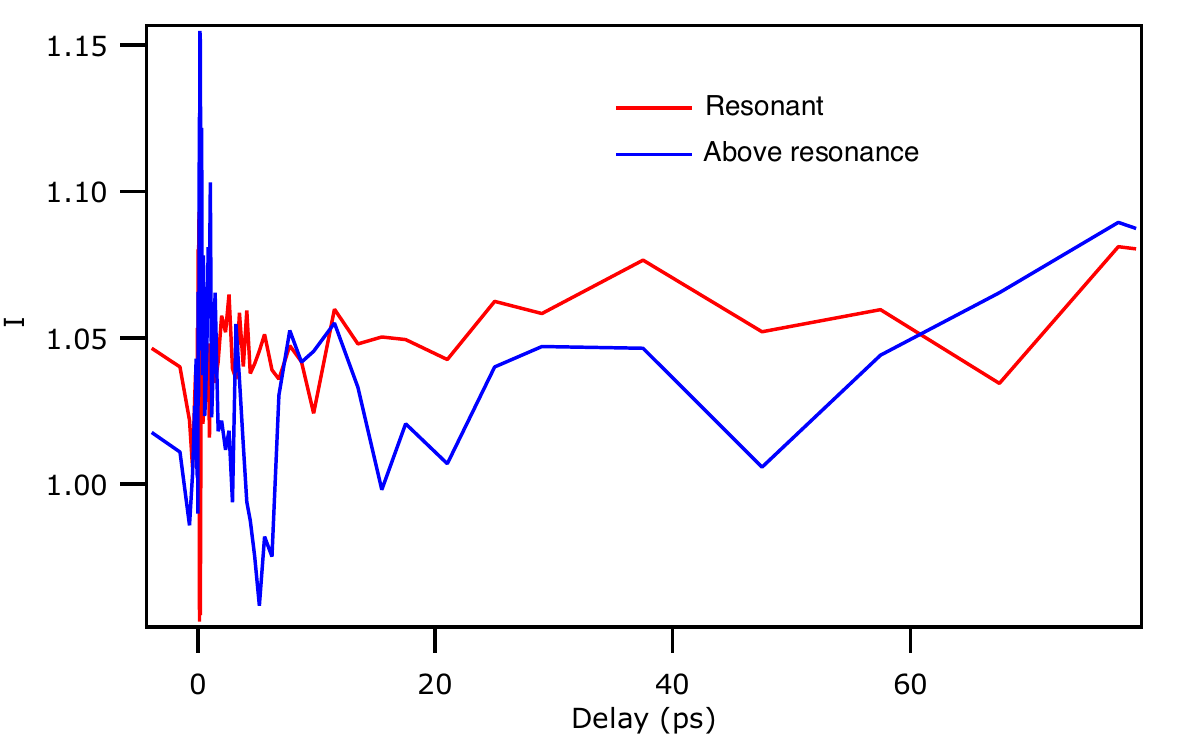}
    \label{Fig_s5_cut}	
    \caption{Time-evolution of the spectral weight $I$ of the fits of tr-RC data of above-resonance (blue) and resonant (red) pumping. Both traces are extracted from the highest initial excitation density of $n_0= 3.1$ $10^{12}$ cm$^{-2}$. The spectral weight does not show any significant dynamics within the noise level.}

	\end{figure}

\section{tr-TC of monolayer WS\textsubscript{2} on fused silica}

	\begin{figure}[ht]
	\centering
	\includegraphics[width=130mm]{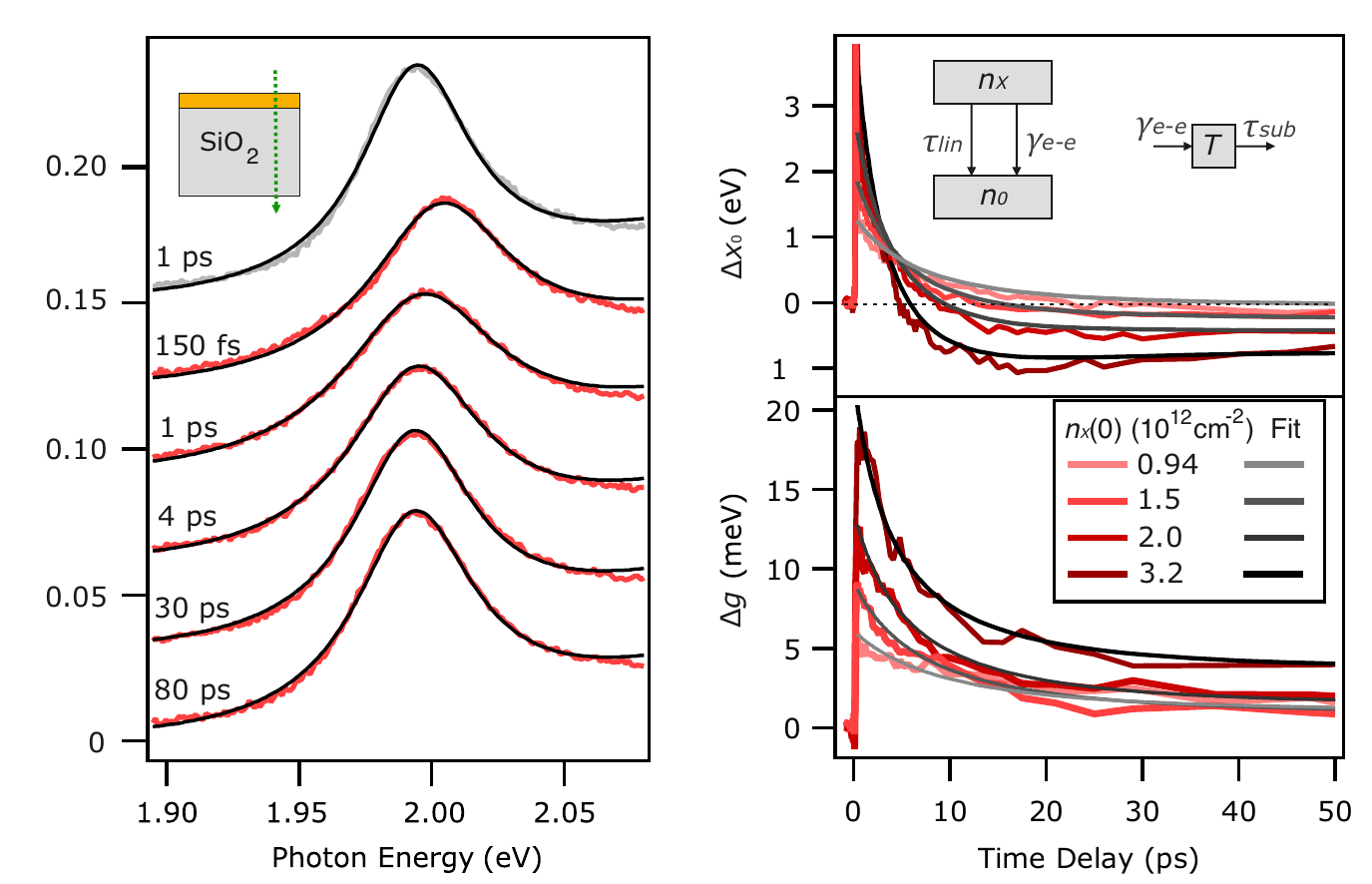}
    	
    \caption{\label{Fig_ss4}a) TC spectra (coloured markers) and respective fits (grey traces) as a function of photon energy for a few selected delays. Traces are vertically shifted for clarity. b) Resonant pumping exciton peak shift ($\Delta x_0$) and linewidth ($\Delta g$) dynamics as a function of time delay for different fluences with corresponding best fits. Inset: Sketch of the two energy level model.}

	\end{figure}
We investigated a monolayer of WS$_2$ placed on a thick fused silica substrate, sketched in the inset of Fig. \ref{Fig_ss4}.a).
We performed time-resolved transmittance contrast (tr-TC) spectroscopy. The excitation conditions are the same as in the experiment described in the main text. We monitor the transmittance contrast $TC= (T_{sub}-T_{sub+WS_2})/T_{sub}$ of the sample as a function of the time delay as described in \cite{Ste:21}.

The TC spectra for a few selected pump-probe delays are reported as a function of the probe photon energy in Fig. \ref{Fig_ss4}.a). The negative time delay TC spectra correspond to the steady-state spectrum in analogy to the RC experiment described in the main text.
The figure reveals that the excitonic resonance undergoes peak shift and broadening upon photoexcitation. We quantify these effects using the analysis method and fit function introduced in our previous work \cite{Ste:21}. We model the complex dielectric function of the exciton with a Lorentzian and derive the analytical expression of the TC spectrum using the 2D linearized model \cite{Heinz_th}. The following expression gives the fit function for the TC:

\begin{equation}
TC=- Re\left[ \frac{2}{1+n_{sub}} Z_0\sigma^s\right]
\tag{S.5}
\end{equation}
\begin{equation}
Z_0\sigma^s=- i2\pi (n_{WS_2}^2-1)(d/\lambda)
\tag{S.6}
\label{eq2}
\end{equation}
Where $n_{sub}$ and $n_{WS_2}$ are the refractive indices of the substrate and monolayer WS$_2$ respectively, $Z_0\sigma^s$ is the dimensionless sheet conductivity, $d$ is the thickness of the monolayer, and $\lambda$ is the wavelength of the light. Here $n_{WS_2}$ is given by the square root of Eq. \ref{lor}, assuming that the exciton contribution to the dielectric function can be described with a Lorentz oscillator.

The time-dependent peak shift $\Delta x_0$ and linewidth $\Delta g$ of the exciton for different fluences are extracted with an analogous procedure to the one described in the main text, and they are reported in Fig. \ref{Fig_ss4}.b).
We fit the time-dependent broadening and peak position with the equation introduced in the main text, and the best fit curves are shown in Fig. \ref{Fig_ss4}.b).
Based on our previous study \cite{Ste:21}, we expect the dynamic response of the exciton to be the same in the two experiments for the same excitation conditions. The extracted parameters are reported in table \ref{tbl1} where they are compared to the best fit parameters extracted in the main text. The extracted parameters of the two experiments coincide, thereby demonstrating the reliability of the model employed to model the photoinduced exciton dynamics.

 \begin{table*} [ht]
\begin{center}

\begin{tabular}{||c | c c c c c c||} 
 \hline

 & $\tau_{lin}$ & $\gamma_{e-e}$ & $\tau_{sub}$ & $\alpha$  &$D$ & $\beta$ \\ [0.5ex] 
 \hline
 Si/SiO$_2$ & 18$\pm$5 & 0.06$\pm$0.01 & 14$\pm$8 & (6.6$\pm$0.6) 10$^{-12}$ & (5.6$\pm$0.2) 10$^{-12}$ & 0.4$\pm$0.1\\ 
 \hline
 FS & 18$\pm$5 & 0.06$\pm$0.01 & 14$\pm$8 & (6.0$\pm$0.4) 10$^{-12}$ & (5.3$\pm$0.3) 10$^{-12}$ & 0.4$\pm$0.1 \\
 \hline
  & ps & cm$^2$/s & ps &cm$^2$  & meV cm$^2$ & meV/K \\
 \hline

\end{tabular}
\end{center}
 \caption{\label{tbl1}Best global fit parameters obtained from the fit with the two energy level model from experiment on layered Si/SiO$_2$ substrate and FS substrate.}

\end{table*}

\section{Calculation of absorption and photoexcited particle density}

	\begin{figure}[ht]
	\centering
	\includegraphics[width=130mm]{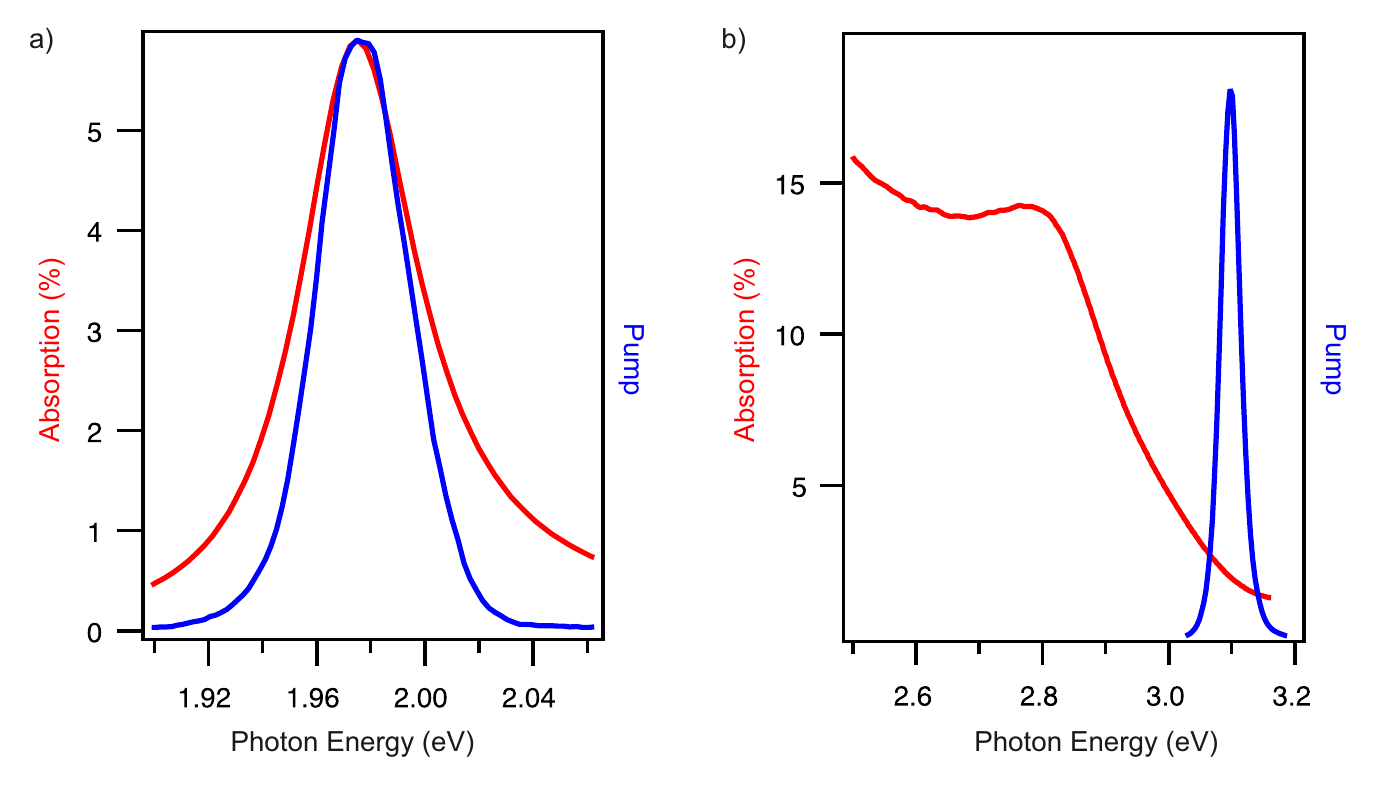}
	\caption{\label{Fig_s3} Selective absorption of monolayer WS$_2$ deposited on a layered Si/SiO$_2$ substrate (red trace) as a function of photon energy for two different energy ranges. Pump spectrum (blue trace) as a function of photon energy.}

	\end{figure}

From the fitting of the steady-state RC spectrum before time zero, we extract the steady-state exciton contribution to the dielectric function, corresponding to a Lorentzian with parameters $x_0=1.973$ eV, $g=54$ meV and $I=1.05$.  

For the monolayer WS$_2$ deposited the Si/SiO$_2$, the calculation of the absorption is complicated due to the layered structure of the sample, as discussed in \cite{Ste:21}. To calculate the absorption, we employed a Matlab script introduced by Jannin et al. in \cite{Alex} based on a transfer matrix formalism for calculating optical observables of complex heterostructures. The dielectric function of the WS$_2$, given by a Lorentzian with the parameters retrieved from the steady-state fitting, was fed to the script. 
The script calculates the monolayer’s selective absorption and returns the absorption spectrum, which is reported in Fig. \ref{Fig_s3}.a) with the red trace.
The normalized pump lineshape is depicted in the same figure (blue trace).

The effective absorption of the WS$_2$ layer is given by the integral of the multiplication of the absorption by the normalized pump’s lineshape and gives a value of 4.6\%. Finally, the absorbed photon density is obtained by multiplying the effective absorption and the incident peak fluence.

Regarding the absorption calculation in the case of the above resonance photoexcitation, the probe does not cover the energy range around 3.1 eV. Therefore, we cannot directly retrieve the complex dielectric function needed to calculate the absorption in that energy range. Instead, we turn then to make use of the fact that the absorption is not particularly sensitive to the dielectric environment due to the lack of exciton in this energy range. Therefore, we employ literature values for the dielectric function \cite{Ansari}, and we feed the latter to the same Matlab script used for the previous calculation of the absorption in the proximity of the excitonic resonance.
The corresponding absorption, for the energy range close to the above resonance pumping, is shown in Fig. \ref{Fig_s3}.b) with the red trace. The effective absorption is calculated in total analogy to the resonant pumping case, returning a value of 1.8\%. The absorbed photon density is obtained with the same procedure as for the resonant pumping.
\pagebreak 
\printbibliography